\documentclass[journal]{IEEEtran}

\usepackage{graphicx} \usepackage{bm} \usepackage{amsmath}
\usepackage{cite}
\usepackage{amssymb} 

\usepackage{amsmath,amsthm,amssymb,mathrsfs}
\usepackage{bm}

\ifCLASSINFOpdf
\else
\fi
\begin{document}

\title{Synchronization of Spin Torque Oscillators through \\ Spin Hall Magnetoresistance}
\author{Tomohiro~Taniguchi
        \\
        National Institute of Advanced Industrial Science and Technology (AIST), 
              Spintronics Research Center, 
              Tsukuba, Ibaraki 305-8568, Japan 
}

\maketitle

\begin{abstract}

Spin torque oscillators placed onto a nonmagnetic heavy metal show 
synchronized auto-oscillations due to the coupling originating from spin Hall magnetoresistance effect. 
Here, we study a system having two spin torque oscillators under the effect of the spin Hall torque, 
and show that switching the external current direction enables us to control the phase difference of the synchronization 
between in-phase and antiphase. 

\end{abstract}


\IEEEpeerreviewmaketitle


\section{Introduction}
\label{sec:Introduction}

\IEEEPARstart{S}{ubstantial} efforts, such as material investigation, 
structural improvement, and theoretical analysis, have been made to develop 
high-performance spin torque oscillators 
because of its potential to be applied in microwave generators and magnetic recording head [1-7]. 
In particular, synchronization of spin torque oscillators is becoming an exciting topic 
because of the possibility in enhancing emission power 
and extending the technology to new practical applications such as bio-inspired computing [8-10]. 
Several methods have been proposed and demonstrated 
to stabilize the synchronizations, based on spin wave propagation [11,12], electric current injection [13,14], mediation of antivortex [15], 
microwave field [16], stochastic noise in current [17], or dipole coupling [18]. 


The spin torque oscillator based on the spin Hall effect [19,20] has been developed recently. 
It has the advantage of easier fabrication and 
that it is unnecessary to apply electric current to the ferromagnet directly. 
A stable auto-oscillation of the magnetization around the in-plane easy axis induced by the spin Hall effect 
was observed in CoFeB/Ta heterostructure in 2012 [21]. 
The synchronization of the spin torque oscillators in the spin Hall geometry 
induced by the spin wave propagation was also demonstrated [22]. 


The spin Hall effect causes another interesting phenomenon related to magnetoresistance effect. 
It was recently found that the resistance of ferromagnetic/nonmagnetic bilayers depends on the magnetization direction in the ferromagnet, 
even when the ferromagnet is an insulator [23-26]. 
This new type of magnetoresistance effect, named as spin Hall magnetoresistance, 
originates from additional electric currents generated by 
the charge-spin conversion due to the direct and inverse spin Hall effects [27]. 
The spin Hall magnetoresistance has been confirmed by measuring the longitudinal and transverse voltages. 
Let us imagine that another ferromagnet is placed onto the nonmagnetic heavy metal in the longitudinal or transverse direction. 
The electric current generated through the spin Hall magnetoresistance effect will be injected into this second ferromagnet 
as spin current by the spin Hall effect, 
and excite the spin torque on the magnetization, and vice versa. 
As a result, the coupled motions of the magnetizations in the ferromagnets are expected. 
This coupling is unavoidable whenever several ferromagnets are placed onto the same nonmagnet. 
Recently, we have confirmed a tangible synchronization of magnetizations having several different phase differences, depending on the material parameters, 
by numerically solving the Landau-Lifshitz-Gilbert (LLG) equation [28]. 


In this paper, the synchronization of spin torque oscillators through the spin Hall magnetoresistance effect is studied theoretically. 
In particular, the control of the phase difference between the oscillators is investigated. 
Considering a system having two oscillators, it is shown that 
switching the direction of the external current applied in one oscillator enables us to control the phase difference 
between in-phase and antiphase. 


This paper is organized as follows. 
In Sec. II, we show the LLG equation including the torque related to the spin Hall magnetoresistance. 
In Sec. III, we compare the synchronized auto-oscillations in two-spin torque oscillators 
for the external electric currents flowing in the same and opposite directions. 
The time necessary to synchronize the oscillators is discussed in Sec. IV. 
The role of the coupling for the oscillators having different anisotropies is discussed in Sec. V. 
The conclusion is summarized in Sec. VI. 


\begin{figure}
\centerline{\includegraphics[width=1.0\columnwidth]{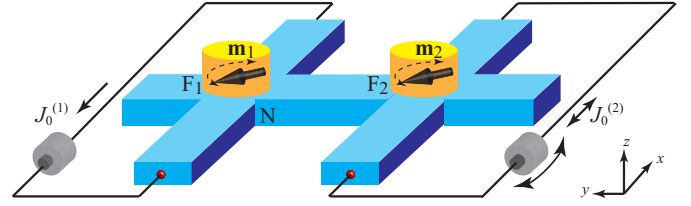}}
\caption{
Schematic view of system under consideration. 
An external voltages are applied along the $x$ direction, generating electric currents in the nonmagnet, N. 
Two ferromagnets, F${}_{1}$ and F${}_{2}$, are aligned along the $y$ direction. 
\vspace{-3.0ex}}
\label{fig:fig1}
\end{figure}



\section{LLG equation}

Figure \ref{fig:fig1} schematically shows the system in this study, 
which consists of two ferromagnets F${}_{k}$ ($k=1,2$) placed onto a nonmagnet, N. 
The ferromagnets are aligned along the $y$ direction. 
External voltages are applied along the $x$ direction, generating the external electric current densities $J_{0}^{(k)}$ 
passing under the F${}_{k}$ layer. 
The directions of the external currents are independently controllable. 
We assume that the ferromagnets are placed onto the nonmagnet at the center along the $x$ direction, 
the electric field in the nonmagnet along the $x$ direction is uniform, 
and the current magnitudes, $|J_{0}^{(1)}|$ and $|J_{0}^{(2)}|$, are the same. 
Therefore, if the spin Hall effect is absent, the electric potentials near the F${}_{1}$/N and F${}_{2}$/N interfaces are the same, 
and electric current does not flow between these interfaces. 
In the presence of the spin-orbit interaction, on the other hand, 
the charge-spin conversion by the direct and inverse spin Hall effects near the F${}_{k}$/N interface gives 
an additional electric current density flowing in the $y$ direction, 
\begin{equation}
  J_{\rm c}^{{\rm F}_{k}/{\rm N}}
  =
  -\left(
    \chi
    m_{k x}
    m_{k y}
    +
    \chi^{\prime}
    m_{k z}
  \right)
  J_{0}^{(k)},
  \label{eq:current_SMR}
\end{equation}
where $\mathbf{m}_{k}=(m_{k x},m_{k y},m_{k z})$ is the unit vector pointing in the magnetization direction in the F${}_{k}$ layer. 
The dimensionless coefficients, $\chi$ and $\chi^{\prime}$, in metallic ferromagnetic/nonmagnetic bilayers are given by [27,29,30]
\begin{equation}
  \chi
  =
  \frac{\vartheta^{2} \ell_{\rm N}}{d_{\rm N}}
  \left[
    {\rm Re}
    \frac{g^{\uparrow\downarrow}}{g_{\rm N} + g^{\uparrow\downarrow} \coth(d_{\rm N}/\ell_{\rm N})}
    -
    \frac{g^{*}}{g_{\rm N}}
  \right]
  \tanh^{2}
  \left(
    \frac{d_{\rm N}}{2 \ell_{\rm N}}
  \right),
\end{equation}
\begin{equation}
  \chi^{\prime}
  =
  -\frac{\vartheta^{2} \ell_{\rm N}}{d_{\rm N}}
  {\rm Im}
  \frac{g^{\uparrow\downarrow}}{g_{\rm N} + g^{\uparrow\downarrow} \coth(d_{\rm N}/\ell_{\rm N})}
  \tanh^{2}
  \left(
    \frac{d_{\rm N}}{2 \ell_{\rm N}}
  \right),
\end{equation}
where $\vartheta$, $\ell_{\rm N}$, and $d_{\rm N}$ are the spin Hall angle, spin diffusion length, and thickness of the nonmagnet, respectively. 
The dimensionless mixing conductance is denoted as $g^{\uparrow\downarrow}$ [31,32], 
whereas $g_{\rm N}/S=h \sigma_{\rm N}/(2e^{2} \ell_{\rm N})$ with the cross section area of the ferromagnetic/nonmagnetic interface, $S$, 
and the conductivity of the nonmagnet, $\sigma_{\rm N}$. 
We introduce $g^{*}$ as 
$1/g^{*}=\{2/[(1-p_{g}^{2})g]\}+\{1/[g_{\rm F} \tanh(d_{\rm F}/\ell_{\rm F})]\}+\{1/[g_{\rm N} \tanh(d_{\rm N}/\ell_{\rm N})]\}$, 
where $g$ and $p_{g}$ are the dimensionless interface conductance and its spin polarization, 
whereas $g_{\rm F}/S=h(1-p_{\sigma}^{2})\sigma_{\rm F}/(2e^{2}\ell_{\rm F})$ 
with the spin diffusion length $\ell_{\rm F}$, the conductivity, $\sigma_{\rm F}$, and its spin polarization, $p_{\sigma}$, of the ferromagnet. 
The thickness of the ferromagnet is $d_{\rm F}$. 
The interface resistance is related to $g$ via $g/S=(h/e^{2})/r$. 
Using the values of the parameters estimated from the experiments in CoFeB/W heterostructure [29] and the first principles calculation [32], 
$\rho_{\rm F}=1/\sigma_{\rm F}=1.6$ k$\Omega$nm $p_{\sigma}=0.72$, $\ell_{\rm F}=1.0$ nm, $\rho_{\rm N}=1.25$ k$\Omega$nm, $\ell_{\rm N}=1.2$ nm, 
$r=0.25$ k$\Omega$nm${}^{2}$, $p_{\rm g}=0.50$, ${\rm Re}[g^{\uparrow\downarrow}/S]=25.0$ nm${}^{-2}$, ${\rm Im}[g^{\uparrow\downarrow}/S]=1.0$ nm${}^{-2}$, $d_{\rm F}=2$ nm, $d_{\rm N}=3$ nm, and $\vartheta=0.27$, 
we set $\chi \simeq 0.01$ and $\chi^{\prime} \simeq -0.0002$. 


The external electric current $J_{0}^{(k)}$ is converted to spin current near the F${}_{k}$/N interface by the direct spin Hall effect, 
and excites the spin torque on the magnetization, $\mathbf{m}_{k}$. 
In addition, in the present system having two ferromagnets, 
the electric current density $J_{\rm c}^{{\rm F}_{k^{\prime}}/{\rm N}}$ 
originated near the F${}_{k^{\prime}}$/N interface through the spin Hall magnetoresistance effect 
and moved to the other ferromagnetic/nonmagnetic interface, F${}_{k}$/N ($k \neq k^{\prime}$), 
is converted to spin current again by the direct spin Hall effect, 
and excites the spin torque on the magnetization, $\mathbf{m}_{k}$. 
Therefore, the LLG equation of the magnetization is [28]
\begin{equation}
\begin{split}
  \frac{d \mathbf{m}_{k}}{dt}
  =&
  -\gamma
  \mathbf{m}_{k}
  \times
  \mathbf{H}_{k}
  +
  \alpha
  \mathbf{m}_{k}
  \times
  \frac{d \mathbf{m}_{k}}{dt}
\\
  &-
  \frac{\gamma \hbar \vartheta_{\rm R} J_{0}^{(k)}}{2eMd_{\rm F}}
  \mathbf{m}_{k}
  \times
  \left(
    \mathbf{e}_{y}
    \times
    \mathbf{m}_{k}
  \right)
  -
  \frac{\gamma \beta \hbar \vartheta_{\rm R} J_{0}^{(k)}}{2eMd_{\rm F}}
  \mathbf{m}_{k}
  \times
  \mathbf{e}_{y}
\\
  &-
  \frac{\gamma \hbar \vartheta_{\rm R} (\chi m_{k^{\prime}x} m_{k^{\prime}y} + \chi^{\prime} m_{k^{\prime}z}) J_{0}^{(k^{\prime})}}{2eMd_{\rm F}}
  \mathbf{m}_{k}
  \times
  \left(
    \mathbf{e}_{x}
    \times
    \mathbf{m}_{k}
  \right)
\\
  &-
  \frac{\gamma \beta \hbar \vartheta_{\rm R} (\chi m_{k^{\prime}x} m_{k^{\prime}y} + \chi^{\prime} m_{k^{\prime}z}) J_{0}^{(k^{\prime})}}{2eMd_{\rm F}}
  \mathbf{m}_{k}
  \times
  \mathbf{e}_{x},
  \label{eq:LLG}
\end{split}
\end{equation}
where $M$, $\alpha$, and $\gamma$ are the saturation magnetization, the Gilbert damping constant, and gyromagnetic ratio of the ferromagnet, respectively, 
and are assumed as $1500$ emu/c.c., $0.005$, and $1.764 \times 10^{7}$ rad/(Oe s) from the experiments [28,32]. 
We also introduce 
\begin{equation}
  \vartheta_{\rm R(I)}
  =
  \vartheta
  \tanh
  \left(
    \frac{d_{\rm F}}{2 \ell_{\rm F}}
  \right)
  {\rm Re}
  \left(
    {\rm Im}
  \right)
  \frac{g^{\uparrow\downarrow}}{g_{\rm N} + g^{\uparrow\downarrow} \coth(d_{\rm N}/\ell_{\rm N})},
\end{equation}
and $\beta=-\vartheta_{\rm I}/\vartheta_{\rm R}$, 
which are also estimated from the parameters written above as $\vartheta_{\rm R} \simeq 0.167$ and $\beta \simeq -0.01$. 


The first and second terms on the right hand side of Eq. (\ref{eq:LLG}) represent 
the torque due to the magnetic field $\mathbf{H}_{k}$ and the Gilbert damping torque. 
The third and fourth terms are the conventional spin Hall torques 
corresponding to the anti-damping (or Slonczewski-like [34]) torque and the field-like torque, respectively. 
Since the auto-oscillation of the magnetization by this spin Hall torque was observed in the in-plane magnetized ferromagnet [21], 
we assume that the magnetic field consists of an in-plane anisotropy field $H_{\rm K}$ along the $y$ direction 
and a shape anisotropy field along the $z$ direction as 
\begin{equation}
  \mathbf{H}_{k}
  =
  H_{\rm K}
  m_{k y}
  \mathbf{e}_{y}
  -
  4 \pi M 
  m_{k z}
  \mathbf{e}_{z}, 
\end{equation}
where $H_{\rm K}$ is set as $200$ Oe. 
Note that there are two stable states of the magnetization, $\mathbf{m}_{k} = \pm \mathbf{e}_{y}$. 
It is known that the auto-oscillation around the $y$ axis is excited when the external current density is larger than a critical value [35], 
\begin{equation}
  J_{\rm c}
  =
  \pm
  \frac{2 \alpha eMd_{\rm F}}{\hbar \vartheta_{\rm R}}
  \left(
    H_{\rm K}
    +
    2\pi M
  \right), 
  \label{eq:critical_current}
\end{equation}
where the double sign means that the upper (lower) when the initial state is near $\mathbf{m}_{k}=+(-)\mathbf{e}_{y}$. 
On the other hand, the last two terms in Eq. (\ref{eq:LLG}) originate from the electric current given by Eq. (\ref{eq:current_SMR}) 
contributing to the spin Hall magnetoresistance effect. 
These coupling torques are on the order of $\vartheta^{3}$, whereas the conventional spin Hall torque is proportional to $\vartheta$. 
Thus, the coupling torque is at least two orders of magnitude smaller than the conventional torque. 
Nevertheless, these torques result in coupled motions of the magnetizations in the F${}_{1}$ and F${}_{2}$ layers. 


The right hand side of Eq. (\ref{eq:LLG}) becomes zero when the magnetization points to the $y$ direction. 
In the following, small deviations of $\mathbf{m}_{1}$ and $\mathbf{m}_{2}$ from the $y$ axis are assumed as 
$\mathbf{m}_{1}=(\cos 80^{\circ},\sin 80^{\circ},0)$ and $\mathbf{m}_{2}=(\cos 95^{\circ},\sin 95^{\circ},0)$ 
at the initial state, 
except Fig. \ref{fig:fig4} where the dependences of the time necessary to synchronize the oscillations on the initial conditions is investigated. 


\section{Synchronization of auto-oscillations}

A key quantity in the synchronization of oscillators is the phase difference. 
Its precise control is of interest for both nonlinear science and practical applications [9,10]. 
For example, in the theoretical study of the self-synchronization by the delayed feedback [36], 
the phase difference between a vortex oscillator and the feedback current is controlled by the delay time. 
In the present system based on the spin Hall effect, it was found that the phase difference can be varied 
when the value of the field-like torque strength, $\beta$, is altered [28]. 
However, $\beta$ is determined by the material parameters, and once a sample is fabricated experimentally, 
its value, as well as sign, cannot be altered. 
Therefore, for practical studies, alternative proposal to control the phase difference will be necessary. 
We note here that the direction and magnitude of the external currents in Ref. [28] were assumed to be identical among the oscillators.

One way proposed here is to control the phase difference between the oscillators by 
reversing the direction of the external electric current. 
Namely, for example, the electric current under the F${}_{1}$ layer always flows in the positive $x$ direction ($J_{0}^{(1)}>0$), 
whereas that under the F${}_{2}$ layer flows in either the positive ($J_{0}^{(2)}>0$) or negative ($J_{0}^{(2)}<0$) $x$ direction. 
We note that the current $J_{0}^{(2)}$ is a direct current, and our proposal does not mean that an alternative current is applied. 
When $J_{0}^{(2)}$ is also positive, the phase difference is zero, i.e., the in-phase synchronization is realized [28]. 
On the other hand, when $J_{0}^{(2)}$ is negative, the phase difference becomes antiphase, as shown below. 


Before discussing the phase difference, 
we note that the magnetization direction should be changed to 
the appropriate direction of $\mathbf{m}_{k} \simeq \pm \mathbf{e}_{y}$ 
before exciting the auto-oscillation. 
In our definition, positive (negative) external current $J_{0}^{(k)}$ excites the auto-oscillation of 
the magnetization $\mathbf{m}_{k}$ around the positive (negative) $y$ direction, 
according to Eq. (\ref{eq:critical_current}). 
Since $J_{0}^{(1)}$ is always positive for the convention, as mentioned above, 
the initial state of $\mathbf{m}_{1}$ should be close to the positive $y$ direction. 
On the other hand, the initial state of $\mathbf{m}_{2}$ should be close to the positive (negative) $y$ direction 
when $J_{0}^{(2)}$ is positive (negative). 
Note that the magnetization direction in the F${}_{2}$ layer can be reversed between the positive and negative $y$ directions 
when $J_{0}^{(2)}$ exceeds another critical value [37], 
\begin{equation}
  J^{*}
  =
  \pm
  \frac{4 \alpha eMd_{\rm F}}{\hbar \vartheta_{\rm R}}
  \sqrt{
    4\pi M 
    \left(
      H_{\rm K}
      +
      4\pi M
    \right)
  }.
  \label{eq:switching_current}
\end{equation}

Then, let us explain the reason why we consider that the phase difference can be altered 
by changing the direction of the external electric current under the F${}_{2}$ layer. 
The magnetization in the F${}_{1}$ layer oscillates around the $y$ axis with the counterclockwise direction. 
When the magnetization in the F${}_{2}$ layer also points to the positive $y$ direction 
and oscillates by the positive current, its precession direction is also counterclockwise. 
On the other hand, when $\mathbf{m}_{2}$ oscillates around the negative $y$ direction by the negative current, 
its precession direction is clockwise. 
Therefore, the complete in-phase synchronization between $\mathbf{m}_{1}$ and $\mathbf{m}_{2}$ is no longer possible. 
Note that the coupling torques include the term 
$(\chi m_{k^{\prime} x} m_{k^{\prime} y} + \chi^{\prime} m_{k^{\prime} z})J_{0}^{(k^{\prime})}$, 
as can be seen from Eq. (\ref{eq:LLG}). 
This term determines the phase difference between the spin torque oscillators. 
For example, the dynamics of $\mathbf{m}_{1}$ is affected by the term 
$(\chi m_{2 x} m_{2 y} + \chi^{\prime} m_{2 z})J_{0}^{(2)}$, 
which will be an even function by the reversal of the current direction. 
We notice that the term $m_{2 y} J_{0}^{(2)}$ in the coupling torque does not change the sign 
by reversing the direction of $J_{0}^{(2)}$. 
Then, it is expected from the term $m_{2 x} m_{2 y} J_{0}^{(2)}$ in the coupling torque 
that the phase difference between $m_{1x}$ and $m_{2x}$ does not change by reversing the direction of $J_{0}^{(2)}$. 
On the other hand, the phase difference between $m_{1z}$ and $m_{2z}$ will be altered by changing the current direction 
because the coupling torque includes the term $m_{2 z} J_{0}^{(2)}$. 


\begin{figure}
\centerline{\includegraphics[width=1.0\columnwidth]{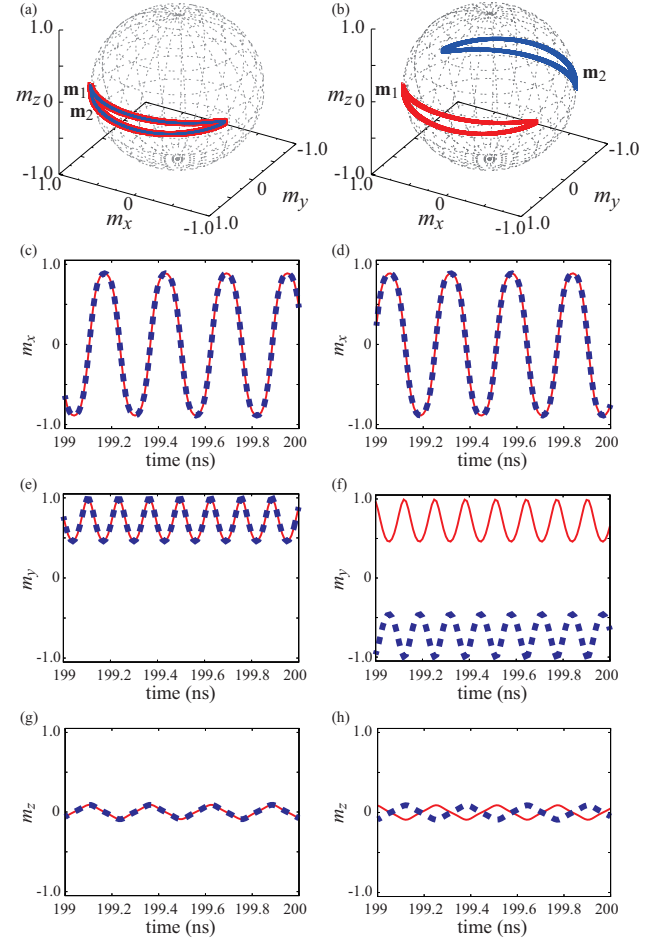}}
\caption{
Dynamic orbits of the magnetizations in the F${}_{1}$ (red) and F${}_{2}$ (blue) layers in the synchronized auto-oscillation states are shown in (a) and (b). 
The external current densities are $(J_{0}^{(1)},J_{0}^{(2)})=(+30,+30) \times 10^{6}$ A/cm${}^{2}$ in (a) 
and $(J_{0}^{(1)},J_{0}^{(2)})=(+30,-30) \times 10^{6}$ A/cm${}^{2}$ in (b). 
The time evolutions of the $x$, $y$, and $z$ components of $\mathbf{m}_{1}(t)$ (red, solid) and $\mathbf{m}_{2}(t)$ (blue, dashed) 
for $(J_{0}^{(1)},J_{0}^{(2)})=(+30,+30) \times 10^{6}$ A/cm${}^{2}$ are shown in (c), (e), and (g), respectively, 
whereas those for $(J_{0}^{(1)},J_{0}^{(2)})=(+30,-30) \times 10^{6}$ A/cm${}^{2}$ are shown in (d), (f), and (h), respectively. 
\vspace{-3.0ex}}
\label{fig:fig2}
\end{figure}



To confirm the validity of these considerations, 
now let us investigate the oscillation behaviors of the magnetizations for positive and negative $J_{0}^{(2)}$ 
by numerically solving Eq. (\ref{eq:LLG}). 
Figures \ref{fig:fig2}(a) and \ref{fig:fig2}(b) show the orbits of the auto-oscillations 
of $\mathbf{m}_{1}(t)$ (red) and $\mathbf{m}_{2}(t)$ (blue) for the external current densities 
(a) in the same directions, $(J_{0}^{(1)},J_{0}^{(2)})=(+30,+30) \times 10^{6}$ A/cm${}^{2}$, 
and (b) in the opposite directions, $(J_{0}^{(1)},J_{0}^{(2)})=(+30,-30) \times 10^{6}$ A/cm${}^{2}$. 
The time evolutions of the $x$, $y$, and $z$ components of $\mathbf{m}_{1}(t)$ (red solid line) and $\mathbf{m}_{2}(t)$ (blue dashed line) 
are shown in Figs. \ref{fig:fig2}(c), \ref{fig:fig2}(e), and \ref{fig:fig2}(g), respectively, 
where the external electric current flow in the same directions, 
The in-phase synchronization, $\mathbf{m}_{1}(t)=\mathbf{m}_{2}(t)$, is confirmed from these figures, as reported in Ref. [28]. 
On the other hand, when the external electric currents flow in the opposite directions, 
the magnetization in the F${}_{1}$ layer oscillates around the positive $y$ direction, 
whereas that in the F${}_{2}$ layer oscillates in the negative $y$ direction, as shown in Figs. \ref{fig:fig2}(b) and \ref{fig:fig2}(f). 
The $x$ components of the magnetizations are still synchronized in in-phase state, 
but the $z$ components are synchronized in antiphase state, i.e., $m_{1x}(t)=m_{2x}(t)$ and $m_{1z}(t)=-m_{2z}(t)$, 
as shown in Figs. \ref{fig:fig2}(d) and \ref{fig:fig2}(h), respectively. 
These results are consistent with the above expectations. 



\begin{figure}
\centerline{\includegraphics[width=1.0\columnwidth]{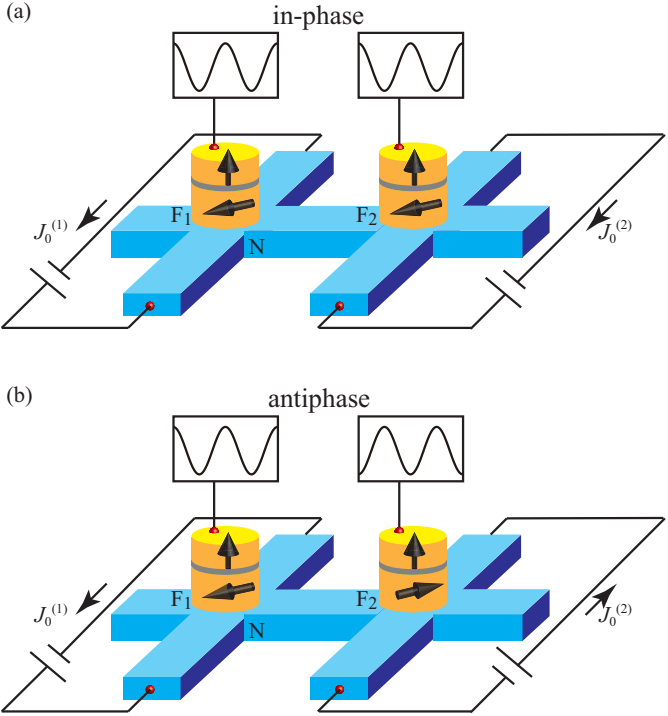}}
\caption{
Schematic views of the system to measure (a) the in-phase and (b) antiphase synchronization of the spin torque oscillators. 
In addition to the structure in Fig. \ref{fig:fig1}, 
perpendicularly magnetized ferromagnets are placed on to the spin torque oscillators 
to detect the oscillations of the magnetizations. 
\vspace{-3.0ex}}
\label{fig:fig3}
\end{figure}


The results shown in Figs. \ref{fig:fig2}(g) and \ref{fig:fig2}(h) indicate a possibility 
to obtain the synchronized signals from the spin torque oscillators 
with both in-phase and antiphase, depending on the direction of the external electric currents. 
A method to measure such signals is to add giant magnetoresistive (GMR) 
or tunneling magnetoresistance (TMR) structures 
on the oscillators, as done in the experiments [18], 
and measure the oscillating resistances through the GMR or TMR effect. 
The magnetization directions of the ferromagnets added on the oscillators should point to the $z$ direction 
because the $z$ components of the magnetizations in the oscillators show either in-phase or antiphase synchronization, 
as shown in Figs. \ref{fig:fig2}(g) and \ref{fig:fig2}(h). 
Figures \ref{fig:fig3}(a) and \ref{fig:fig3}(b) show schematic views of such structures, 
where (a) the in-phase oscillating signals are obtained when both $J_{0}^{(1)}$ and $J_{0}^{(2)}$ flow in the same directions, 
whereas (b) the antiphase signals are obtained when $J_{0}^{(1)}$ and $J_{0}^{(2)}$ flow in the opposite directions.


\section{Time necessary to synchronize oscillators}

As mentioned above, the results shown in Fig. \ref{fig:fig2} are obtained from the LLG equation 
with the initial conditions of $\mathbf{m}_{1}=(\cos 80^{\circ},\sin 80^{\circ},0)$ and $\mathbf{m}_{2}=(\cos 95^{\circ},\sin 95^{\circ},0)$. 
One might be interested in how the initial conditions of the magnetizations affect this conclusion, 
In this section, let us discuss the role of the initial state on the synchronization. 


The oscillators in the present model have two energetically stable state at $\mathbf{m}_{k}=\pm\mathbf{e}_{y}$, 
and these two states are separated by the energy barrier. 
The phase difference in the synchronized state is zero when two magnetizations initially stay near the same stable direction, 
whereas it becomes antiphase when they stay near the different directions at the initial state. 
In other words, the phase difference is zero when $m_{1y}(t=0)/m_{2y}(t=0)>0$, 
whereas it is antiphase when $m_{1y}(t=0)/m_{2y}(t=0)<0$. 
This conclusion is not affected by the specific values of $m_{1y}(t=0)$ and $m_{2y}(t=0)$. 


On the other hand, the time necessary to synchronize the oscillators depends on the initial states of the magnetizations. 
We investigate the relation between the time and phase difference for several initial states from the numerical simulations. 
The initial states of the magnetizations are distributed by the thermal fluctuations. 
Due to the large demagnetization field along the $z$ direction, 
the magnetizations lie almost in the $xy$ plane with the averaged angle from the $y$ axis to the $x$ axis as 
\begin{equation}
  \langle \theta \rangle 
  =
  \frac{\int_{0}^{\pi/2} \theta \exp(\Delta_{0}\cos^{2}\theta) \sin\theta d \theta}{Z},
  \label{eq:angle_distribution}
\end{equation}
where $\Delta_{0}=MH_{\rm K}V/(2k_{\rm B}T)$ with the volume $V$, the Boltzmann constant $k_{\rm B}$, and the temperature $T$ is 
the thermal stability, whereas $Z$ is the partition function defined as 
\begin{equation}
  Z
  =
  \int_{0}^{\pi/2} 
  \exp
  \left(
    \Delta_{0}
    \cos^{2}\theta
  \right) 
  \sin\theta 
  d \theta.
  \label{eq:partition_function}
\end{equation}
Since the exponentials in Eqs. (\ref{eq:angle_distribution}) and (\ref{eq:partition_function}) are dominated near $\theta \simeq 0$, 
$\langle \theta \rangle$ becomes 
\begin{equation}
\begin{split}
  \langle \theta \rangle 
  &\simeq
  \frac{\int_{0}^{\infty} \theta^{2} \exp[\Delta_{0}(1-\theta^{2})] d \theta}{\int_{0}^{\infty} \theta \exp[\Delta_{0}(1-\theta^{2})] d \theta}
\\
  &=
  \frac{1}{2}
  \sqrt{
    \frac{\pi}{\Delta_{0}}
  }.
\end{split}
\end{equation}
Assuming $\Delta_{0}=60$, which is a required value for memory applications [38], 
$\langle \theta \rangle \simeq 6.6^{\circ}$. 
We consider that the relative angle between the magnetizations affects the time necessary to synchronize the oscillators. 
Therefore, we change the values of $\varphi_{1}$ in $\mathbf{m}_{1}(0)=(\cos\varphi_{1},\sin\varphi_{1},0)$ 
around $\varphi_{1}=\pi/2 \pm \langle \theta \rangle$, while the initial condition of the other magnetization is fixed as 
$\mathbf{m}_{2}(0)=(\cos 95^{\circ},\sin 95^{\circ},0)$.


\begin{figure}
\centerline{\includegraphics[width=1.0\columnwidth]{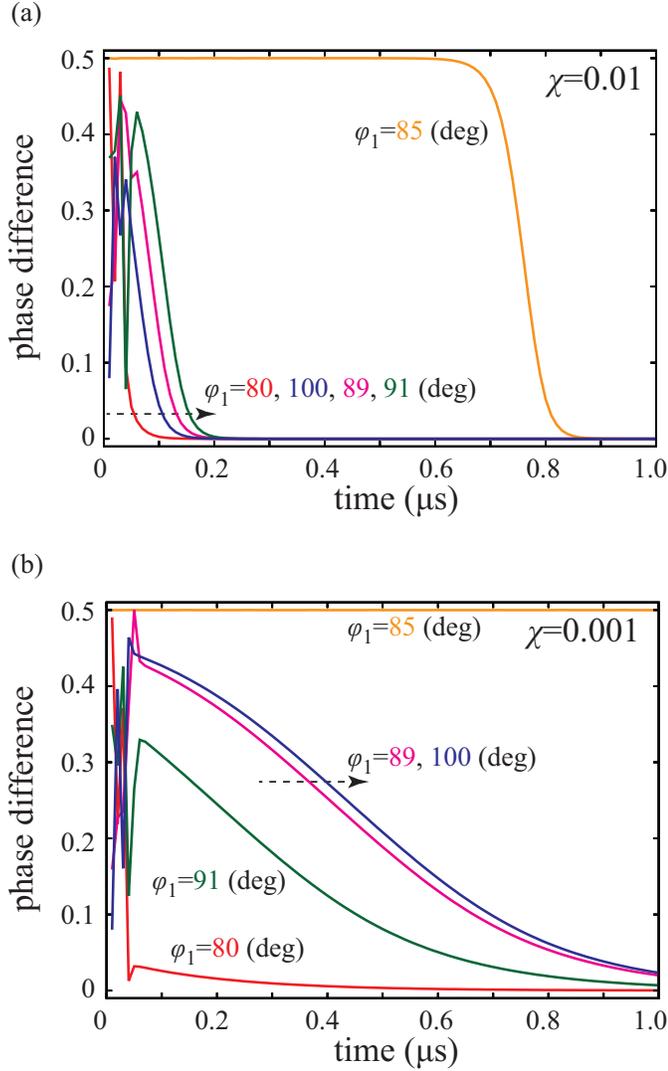}}
\caption{
Time evolutions of the phase difference for several initial conditions of $\varphi_{1}$, 
where the coupling parameter $\chi$ is (a) 0.01 and (b) 0.001. 
\vspace{-3.0ex}}
\label{fig:fig4}
\end{figure}



In Fig. \ref{fig:fig4}(a), we show the time evolutions of the phase difference between the oscillators 
for several initial states, $\varphi_{1}=80^{\circ},81^{\circ},85^{\circ},91^{\circ}$, and $100^{\circ}$. 
We note that we did not calculate it for $\varphi=95^{\circ}$ because, in this case, two magnetizations have the same initial conditions, 
and therefore, the in-phase synchronization is excited from the initial state. 
We consider the case that two currents flow in the same direction. 
For all cases, the phase difference finally becomes zero, i.e., in-phase, as in the results in the previous section. 
The phase difference in the vertical axis is calculated by the same algorithm in Ref. [28], 
where the zero corresponds to the in-phase synchronization whereas 0.50 corresponds to the antiphase synchronization. 
The results shown in Fig. \ref{fig:fig4}(a) indicate that 
the time necessary to excite the synchronization is typically on the order of hundred nanoseconds. 


An exception is found for $\varphi_{1}=85^{\circ}$ in Fig. \ref{fig:fig4}(a), 
where the time almost 1 $\mu$s is necessary to fix the phase difference. 
This result is explained as follows. 
We note that the initial phase difference of the magnetizations for $\varphi_{1}=85^{\circ}$ is exactly the antiphase, 
i.e., $\mathbf{m}_{1}(0)$ is $(\cos 85^{\circ},\sin 85^{\circ},0)$, whereas $\mathbf{m}_{2}(0)=(\cos 95^{\circ},\sin 95^{\circ},0)$. 
As clarified in our previous work [28], 
the present model has fixed points at the in-phase and antiphase synchronized state, 
where the former corresponds to a stable fixed point (an attractor) whereas the latter is the unstable one. 
Since the initial states for $\varphi_{1}=85^{\circ}$ corresponds to the fixed point of the synchronization, 
even though it is unstable, 
a relatively long time is necessary to reach the stable in-phase synchronized state. 


It should be reminded that the above calculations are performed for the coupling constant of $\chi\simeq 0.01$. 
We note that the time necessary to synchronize the oscillators depends on the coupling strength. 
For comparison, we also show the time evolution of the phase difference for $\chi = 0.001$ in Fig. \ref{fig:fig4}(b). 
It is found that the time longer than that shown in Fig. \ref{fig:fig4}(a) is necessary to reach the synchronized state. 
The time to fix the phase difference is typically on the order of 1 $\mu$s, 
except for the initial condition corresponds to the unstable fixed point ($\varphi_{1}=85^{\circ}$), 
where the phase difference does not reach to the in-phase state due to the same reason mentioned above. 



\section{Oscillators having different parameters}

\begin{figure}
\centerline{\includegraphics[width=1.0\columnwidth]{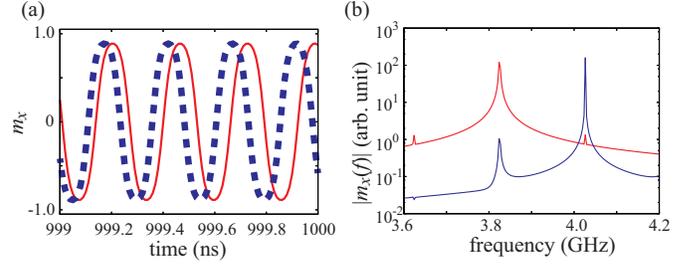}}
\caption{
(a) The time evolutions of $m_{1x}(t)$ (red, solid) and $m_{2x}(t)$, (blue, dashed). 
(b) The Fourier spectra of $m_{1x}$ and $m_{2x}$ in the oscillating state. 
\vspace{-3.0ex}}
\label{fig:fig5}
\end{figure}

A difficulty using the spin Hall effect as a coupling mechanism of the spin torque oscillators is its small strength. 
We note that the above calculations have been performed for identical oscillators. 
When the oscillators have different parameters, a strong strength of the coupling is required to lock the frequencies of the oscillators: 
if the coupling is weak, the synchronization is easily unlocked. 
Figure \ref{fig:fig5}(a) shows an example of a desynchronized state, 
where the in-plane anisotropy field $H_{\rm K}$ in the F${}_{2}$ layer is changed from 200 Oe to 220 Oe, 
whereas the other parameters are kept to those used in Sec. III. 
The Fourier transformations of $m_{1x}(t)$ and $m_{2x}(t)$ are shown in Fig. \ref{fig:fig2}(b). 
It is clearly shown that the main peaks of $|m_{1x}(f)|$ and $m_{2x}(f)|$ appear different frequencies $f$, 
i.e., the peak frequencies are 3.824 and 4.026 GHz for $m_{1x}$ and $m_{2x}$, respectively. 
We change the value of $H_{\rm K}$ in the F${}_{2}$ layer further down to 205 Oe, 
but do not observe clear frequency lockings. 


The above result on desynchronization is also understood from the following consideration. 
The coupling force in the unit of the angular frequency is roughly estimated 
from Eqs. (\ref{eq:LLG}) and (\ref{eq:critical_current}) as 
\begin{equation}
\begin{split}
  \varOmega
  &\sim
  \frac{\gamma \hbar \chi \vartheta_{\rm R} J_{0}}{2eMd_{\rm F}}
\\
  &\sim
  \alpha
  \gamma
  \chi 
  \left(
    H_{\rm K}
    +
    2\pi M 
  \right),
  \label{eq:locking_range}
\end{split}
\end{equation}
which is 8.5 MHz for the present parameters. 
According to the Adler equation [39], 
the frequency difference between the oscillators should be smaller than $\varOmega/(2\pi)$. 
Using the ferromagnetic resonance (FMR) frequency formula, $\gamma \sqrt{H_{\rm K}(H_{\rm K}+4\pi M)}/(2\pi)$, 
the difference of the anisotropy fields between two oscillators, $\delta H_{\rm K}$, 
satisfying $|\gamma \sqrt{(H_{\rm K}+\delta H_{\rm K})(H_{\rm K}+\delta H_{\rm K}+4\pi M)}/(2\pi)-\gamma \sqrt{H_{\rm K}(H_{\rm K}+4\pi M)}/(2\pi)|<\varOmega/(2\pi)$ 
is estimated to be 0.1 Oe. 
Although the nonlinearity of the oscillation broaden the locking range [4], 
it might be still insufficient to excite the synchronization. 


A narrow locking range arises from not only the smallness of the spin Hall effect 
but also the damping parameter $\alpha$ appeared in Eq. (\ref{eq:locking_range}). 
This term comes from the fact that the spin torque compensates with the damping torque in the self-oscillation state, 
and therefore, the current necessary to excite the oscillation is proportional to $\alpha$. 
A small damping constant is preferable to excite the self-oscillation by low currents, 
but makes it difficult to excite the synchronization between the oscillators having different frequency. 


A possible way to overcome these difficulties is utilizing voltage control of magnetic anisotropy [40,41]. 
Adding an MgO barrier on the oscillating layers and applying a direct voltage, 
the magnetic anisotropy, as well as the oscillation frequency, of each ferromagnet can be tuned to a precise value independently. 
A typical value of the perpendicular anisotropy energy modified by the voltage application is about 
200 $\mu$J/m${}^{2}$ in FeCo for $d_{\rm F}=0.68$ nm [41], 
which corresponds to the perpendicular anisotropy field of 400 Oe, assuming $M=1500$ emu/c.c. 
We note that the three-terminal structure of the spin torque oscillators based on the spin Hall effect is 
suitable to control the in-plane current exciting the self-oscillation 
and the perpendicular voltage modifying the anisotropy independently. 
Therefore, the voltage control of magnetic anisotropy will provide an interesting tool for the synchronization 
of the spin torque oscillator based on the three terminal devices. 



\section{Conclusion}

In conclusion, the synchronization of the auto-oscillations in spin torque oscillators spontaneously excited by the spin Hall effect 
was investigated by considering the coupling torques generated from 
the electric currents contributing to the spin Hall magnetoresistance effect. 
It is shown that the phase difference between the magnetizations become 
either in-phase or antiphase, depending on the direction of the external electric current. 
The result indicates that the precise control of the phase difference between spin torque oscillators is possible 
by using the spin Hall effect as a coupling mechanism. 


\section*{Acknowledgment}

The author is grateful to H. Kubota, S. Tsunegi, Y. Shiota, T. Yorozu, S. Tamaru, 
T. Nagasawa, K. Kudo, and Y. Kawamura for valuable discussions. 
The author also expresses gratitude to S. Yuasa, A. Fukushima, K. Yakushiji, 
S. Iba, A. Spiesser, A. Sugihara, T. Kubota, H. Maehara, and A. Emura for their support and encouragement. 
This work was supported by JSPS KAKENHI Grant-in-Aid for Young Scientists (B) 16K17486. 


\ifCLASSOPTIONcaptionsoff
  \newpage
\fi


\begin{thebibliography}{90}

\bibitem{IEEEhowto:kiselev03}
S. I. Kiseleve, J. C. Sankey, I. N. Krivorotov, N. C. Emley, R. J. Schoelkopf, R. A. Buhrman, and D. C. Ralph,
"Microwave oscillations of a nanomagnet driven by a spin-polarized current", 
\emph{Nature}, vol. 425, p.380, 2003. 

\bibitem{IEEEhowto:rippard04}
W. H. Rippard, M. R. Pufall, S. Kaka, S. E. Russek, and T. J. Silva,
"Direct-Current Induced Dynamics in Co${}_{90}$Fe${}_{10}$/Ni${}_{80}$Fe${}_{20}$ Point Contacts", 
\emph{Phys. Rev. Lett.} vol. 92, Art. no. 027201, 2004. 

\bibitem{IEEEhowto:houssameddine07}
D. Houssameddine, U. Ebels, B. Dela\"et, B. Rodmacq, I. Firastrau, F. Ponthenier, M. Brunet, C. Thirion, J. P. Michel, L. Prejbeanu-Buda, M. C. Cyrille, O. Redon, and B. Dieny,
"Spin-torque oscillator using a perpendicular polarizer and a planar free layer", 
\emph{Nat. Mater.} vol. 6, p.447, 2007. 

\bibitem{IEEEhowto:slavin09}
A. Slavin and V. Tiberkevich,
"Nonlinear Auto-Oscillator Theory of Microwave Generation by Spin-Polarized Current"
\emph{IEEE Trans. Magn.} vol. 45, p.1875, 2009. 

\bibitem{IEEEhowto:kubota13}
H. Kubota, K. Yakushiji, A. Fukushima, S. Tamaru, M. Konoto, T. Nozaki, S. Ishibashi, T. Saruya, S. Yuasa, T. Taniguchi, H. Arai, and H. Imamura,
"Spin-Torque Oscillator Based on Magnetic Tunnel Junction with a Perpendicularly Magnetized Free Layer and In-Plane Magnetized Polarizer",
\emph{Appl. Phys. Express} vol. 6, Art. no. 103003, 2013. 

\bibitem{IEEEhowto:tamaru14}
S. Tamaru, H. Kubota, K. Yakushiji, T. Nozaki, M. Konoto, A. Fukushima, H. Imamura, T. Taniguchi, H. Arai, T. Yamaji, and S. Yuasa,
"Bias field angle dependence of the self-oscillation of spin torque oscillators having a perpendicularly magnetized free layer and in-plane magnetized reference layer", 
\emph{Appl. Phys. Express} vol. 7, Art. no. 063005, 2014. 

\bibitem{IEEEhowto:tsunegi14}
S. Tsunegi, T. Taniguchi, H. Kubota, H. Imamura, S. Tamaru, M. Konoto, K. Yakushiji, A. Fukushima, and S. Yuasa,
"Discontinuous frequency drop in spin torque oscillator with a perpendicularly magnetized FeB free layer", 
\emph{Jpn. J. Appl. Phys.} vol. 53, Art. no. 060307, 2014. 

\bibitem{IEEEhowto:locatelli14}
N. Locatelli, V. Cros, and J. Grollier, 
"Spin-torque building blocks", 
\emph{Nat. Mater.} vol. 13, p.11, 2014. 

\bibitem{IEEEhowto:grollier16}
J. Grollier, D. Querlioz, and M. D. Stiles, 
"Spintronic Nanodevices for Bioinspired Computing", 
\emph{Proc. IEEE} vol. 104, p.2024, 2016. 

\bibitem{IEEEhowto:kudo17}
K. Kudo and T. Morie, 
"Self-feedback electrically coupled spin-Hall oscillator array for pattern-matching operation", 
\emph{Appl. Phys. Express} vol. 10, Art. no. 043001, 2017. 

\bibitem{IEEEhowto:kaka05}
S. Kaka, M. R. Pufall, W. H. Rippard, T. J. Silva, S. E. Russek, and J. A. Katine, 
"Mutual phase-locking of microwave spin torque nano-oscillators", 
\emph{Nature} vol. 437, p.389, 2005. 

\bibitem{IEEEhowto:mancoff05}
F. B. Mancoff, N. D. Rizzo, B. N. Engel, and S. Tehrani, 
"Phase-locking in double-point-contact spin-transfer devices", 
\emph{Nature} vol. 437, p.393, 2005. 

\bibitem{IEEEhowto:rippard05}
W. H. Rippard, M. R. Pufall, S. Kaka, T. J. Silva, and S. E. Russek, 
"Injection Locking and Phase Control of Spin Transfer Nano-oscillators", 
\emph{Phys. Rev. Lett.} vol. 95, Art. no. 067203, 2005. 

\bibitem{IEEEhowto:zhou09}
Y. Zhou and J. Akerman,
"Perpendicular spin torque promotes synchronization of magnetic tunnel junction based spin torque oscillators", 
\emph{Appl. Phys. Lett.} vol. 94, Art. no. 112503, 2009. 

\bibitem{IEEEhowto:ruotolo09}
A. Ruotolo, V. Cros, B. Georges, A. Dussaux, J. Grollier, C. Deranlot, R. Guillemet, K. Bouzehouane, S. Fusil, and A. Fert, 
"Phase-locking of magnetic vortices mediated by antivortices", 
\emph{Nat. Nanotechnol.} vol. 4, p.528, 2009. 

\bibitem{IEEEhowto:urazhdin10}
S. Urazhdin, P. Tabor, V. Tiberkevich, and A. Slavin, 
"Fractional Synchronization of Spin-Torque Nano-Oscillators", 
\emph{Phys. Rev. Lett.} vol. 105, Art. no. 104101, 2010. 

\bibitem{IEEEhowto:nakada12}
K. Nakada, S. Yakata, and T. Kimura, 
"Noise-induced synchronization in spin torque nano oscillators", 
\emph{J. Appl. Phys.} vol. 111, Art. no. 07C920, 2012. 

\bibitem{IEEEhowto:locatelli15}
N. Locatelli, A. Hamadeh, F. A. Araujo, A. D. Belanovsky, P. N. Skirdkov, R. Lebrun, V. V. Naletov, K. A. Zvezdin, M. Munoz, J. Grollier, O. Klein, V. Cros, and G. de Loubens, 
"Efficient Synchronization of Dipolarly Coupled Vortex-Based Spin Transfer Nano-Oscillators", 
\emph{Sci. Rep} vol. 5, p.17039, 2015. 

\bibitem{IEEEhowto:dyakonov71}
M. I. Dyakonov and V. I. Perel,
"Current-induced spin orientation of electrons in semiconductors", 
\emph{Phys. Lett. A}, vol.35, p.459, 1971.

\bibitem{IEEEhowto:hirsch99}
J. E. Hirsch,
"Spin Hall Effect", 
\emph{Phys. Rev. Lett.}, vol.83, Art. no. 1834, 1999.

\bibitem{IEEEhowto:liu12}
L. Liu, C.-F. Pai, D. C. Ralph, ad R. A. Buhrman,
"Magnetic Oscillations Driven by the Spin Hall Effect in 3-Terminal Magnetic Tunnel Junction Devices", 
\emph{Phys. Rev. Lett.}, vol.109, Art. no. 186602, 2012.

\bibitem{IEEEhowto:awad16}
A. A. Awad, P. D\"urrenfeld, A. Houshang, M. Dvornik, E. Iacocca, R. K. Dumas, and J. Akerman, 
"Long-range mutual synchronization of spin Hall nano-oscillators", 
\emph{Nat. Phys.} vol. 13, p. 292, 2017. 

\bibitem{IEEEhowto:huang12}
S. Y. Huang, X. Fan, D. Qu, Y. P. Chen, W. G. Wang, J. Wu, T. Y. Chen, J. Q. Xiao, and C. L. Chien, 
"Transport Magnetic Proximity Effects in Platinum", 
\emph{Phys. Rev. Lett.} vol. 109, Art. no. 107204, 2012. 

\bibitem{IEEEhowto:nakayama13}
H. Nakayama, M. Althammer, Y.-T. Chen, K. Uchida, Y. Kajiwara, D. Kikuchi, T. Ohtani, S. Gepr\"ags, M. Opel, S. Takahashi, R. Gross, G. E. W. Bauer, S. T. B. Goennenwein, and E. Saitoh, 
"Spin Hall Magnetoresistance Induced by a Nonequilibrium Proximity Effect", 
\emph{Phys. Rev. Lett.} vol. 110, Art. no. 206601, 2013. 

\bibitem{IEEEhowto:althammer13}
M. Althammer, S. Meyer, H. Nakayama, M. Schreier, S. Althmannshofer, M. Weiler, H. Huebl, S. Gepr\"ags, M. Opel, R. Gross, D. Meier, C. Klewe, T. Kuschel, J.-M. Schmalhorst, G. Reiss, L. Shen, A. Gupta, Y.-T. Chen, G. E. W. Bauer, E. Saitoh, and S. T. B. Goennenwein, 
"Quantitative study of the spin Hall magnetoresistance in ferromagnetic insulator/normal metal hybrids", 
\emph{Phys. Rev. B} vol. 87, Art. no. 224401, 2013. 

\bibitem{IEEEhowot:hahn13}
C. Hahn, G. de Loubens, O. Klein, M. Viret, V. V. Naletov, and J. B. Youssef, 
"Comparative measurements of inverse spin Hall effects and magnetoresistance in YIG/Pt and YIG/Ta", 
\emph{Phys. Rev. B} 87, Art. no. 174417, 2013. 

\bibitem{IEEEhowto:chen13}
Y.-T. Chen, S. Takahashi, H. Nakayama, M. Althammer, S. T. B. Goennenwein, E. Saitoh, and G. E. W. Bauer, 
"Theory of spin Hall magnetoresistance", 
\emph{Phys. Rev. B} vol. 87, Art. no. 144411, 2013. 

\bibitem{IEEEhowto:taniguchi17}
T. Taniguchi, 
"Dynamic coupling of ferromagnets via spin Hall magnetoresistance", 
\emph{Phys. Rev. B} vol. 95, Art. no. 104426, 2017. 

\bibitem{IEEEhowto:kim16}
J. Kim, P. Sheng, S. Takahashi, S. Mitani, and M. Hayashi, 
"Spin Hall Magnetoresistance in Metallic Bilayers", 
\emph{Phys. Rev. Lett.} vol. 116, Art. no. 097201, 2016. 

\bibitem{IEEEhowto:taniguchi16}
T. Taniguchi, 
"Magnetoresistance generated from charge-spin conversion by anomalous Hall effect in metallic ferromagnetic/nonmagnetic bilayers", 
\emph{Phys. Rev. B} vol. 94, Art. no. 174440, 2016. 

\bibitem{IEEEhowto:brataas00}
A. Brataas, Y. V. Nazarov, and G. E. W. Bauer, 
"Finite-Element Theory of Transport in Ferromagnet-Normal Metal Systems", 
\emph{Phys. Rev. Lett.} vol. 84, p.2481, 2000. 

\bibitem{IEEEhowto:zwierzycki05}
M. Zwierzycki, Y. Tserovnyak, P. J. Kelly, A. Brataas, and G. E. W. Bauer, 
"First-principles study of magnetization relaxation enhancement and spin transfer in thin magnetic films", 
\emph{Phys. Rev. B} vol. 71, Art. no. 064420, 2005. 

\bibitem{IEEEhowto:tsunegi14}
S. Tsunegi, H. Kubota, S. Tamaru, K. Yakushiji, M. Konoto, A. Fukushima, T. Taniguchi, H. Arai, H. Imamura, and S. Yuasa, 
"Damping parameter and interfacial perpendicular magnetic anisotropy of FeB nanopillar sandwiched between MgO barrier and cap layers in magnetic tunnel junctions", 
\emph{Appl. Phys. Express} vol. 7, Art. no. 033004, 2014.

\bibitem{IEEEhowto:slonczewski96}
J. C. Slonczewski,
"Current-driven excitation of magnetic multilayers", 
\emph{J. Magn. Magn. Mater.} vol. 159, p.L1, 1996. 

\bibitem{IEEEhowto:grollier03}
J. Grollier, V. Cros, H. Jaffr\'es, A. Hamzic, J. M. George, G. Faini, J. B. Youssef, H. Le Gall, and A. Fert, 
"Field dependence of magnetization reversal by spin transfer", 
\emph{Phys. Rev. B} vol. 67, Art. no. 174402, 2003. 

\bibitem{IEEEhowto:khalsa15}
G. Khalsa, M. D. Stiles, and J. Grollier, 
"Critical current and linewidth reduction in spin-torque nano-oscillators by delayed self-injection", 
\emph{Appl. Phys. Lett.} vol. 106, Art. no. 242402, 2015. 

\bibitem{IEEEhowto:taniguchi13}
T. Taniguchi, Y. Utsumi, M. Marthaler, D. S. Golubev, and H. Imamura,
"Spin torque switching of an in-plane magnetized system in a thermally activated region", 
\emph{Phys. Rev. B}, vol. 87, Art. no. 054406, 2013. 

\bibitem{IEEEhowto:yakushiji08}
K. Yakushiji, S. Yuasa, T. Nagahamaand A. Fukushima, H. Kubota,T. Katayama and K. Ando,
"Spin-Transfer Switching and Thermal Stability in an FePt/Au/FePt Nanopillar Prepared by Alternate Monatomic Layer Deposition",
\emph{Appl. Phys. Express} vol. 1, Art. no. 041302, 2008. 

\bibitem{IEEEhowto:pikovsly03}
A. Pikovsky, M. Rosenblum, and J. Kurths, 
"Synchronization: A universal concept in nonlinear science",
1st ed., Cambridge Uniersity Press, 2003.

\bibitem{IEEEhowto:maruyama09}
T. Maruyama, Y. Shiota, T. Nozaki, K. Ohta, N. Toda, M. Mizuguchi, A. A. Tulapurkar, T. Shinjo, M. Shiraishi, S. Mizukami, Y. Ando, and Y. Suzuki,
"Large voltage-induced magnetic anisotropy change in a few atomic layers of iron",
\emph{Nat. Nanotechnol.} vol. 4, p.158, 2009. 

\bibitem{IEEEhowto:shiota11}
Y. Shiota, S. Murakami, F. Bonell, T. Nozaki, T. Shinjo, and Y. Suzuki, 
"Quantitative Evaluation of Voltage-Induced Magnetic Anisotropy Change by Magnetoresistance Measurement",
\emph{Appl. Phys. Express} vol. 4, Art. no. 043005, 2011. 

\end{thebibliography}


\end{document}